\documentclass[american,submission,copyright,creativecommons]{eptcs}
\providecommand{\event}{PROLE 2014}

\usepackage[nodate]{datetime}
\usepackage{babel}
\usepackage{amsmath}
\usepackage{mathtools}[2011/02/12]
\usepackage{amsfonts}
\usepackage{amssymb}
\let\vec\relax
\usepackage{MnSymbol}
\usepackage[applemac]{inputenc}
\usepackage{multicol}
\usepackage{tikz} 
\usetikzlibrary{arrows,automata}
\newenvironment{Draftpleasenote*}[1][??]{\begin{draftpleasenote*}[{#1}]}{\end{draftpleasenote*}}

\usepackage[fancyproofs,noextended,squareitemtag]{theorems}[2013/01/24]

\title{Modeling Hybrid Systems in the Concurrent Constraint Paradigm}
\author{
Dami\'an Adalid \qquad Mar\'ia del Mar Gallardo \qquad Laura Titolo
\institute{Dept. Lenguajes y Ciencias de la Computaci\'on\\
E.T.S.I. Inform\'atica \quad
University of M\'alaga\thanks{This work has been
  supported by the Andalusian Excellence Project P11-TIC7659 and the Spanish Ministry of  Economy and Competitiveness project TIN2012-35669 }}
\email{[damian,gallardo,laura.titolo]@lcc.uma.es }
}
\def\authorrunning{Adalid, Gallardo \& Titolo}
\def\titlerunning{Modeling Hybrid Systems in the Concurrent Constraint Paradigm}

\newcommand*{\dfn}{\coloneq}
\newcommand*{\seq}[2][n]{#2_{1}, \dots, #2_{#1}}
\providecommand*{\mathoper}[1]{\mathop{\mathit{#1}}\nolimits}
\providecommand*{\pair}[2]{{\langle #1, \, #2 \rangle}}
\providecommand*{\triple}[3]{{\langle #1, \, #2, \, #3 \rangle}}
\providecommand*{\eg}   {e.g.} 
\providecommand*{\ie}   {i.e.,} 
\providecommand*{\resp} {respectively}

\providecommand*{\wrt}  {w.r.t.}
\makeatletter
\providecommand{\ifempty}[3]{\def\@@@temp{#1}\ifx\@@@temp\@empty#2\else#3\fi}
\makeatother
\providecommand*{\wgmWarn}[1]{\typeout{*** using WgMacros def of \string#1}}

\providecommand*{\parensmathoper}[2]{\ensuremath{\mathoper{#1}\ifempty{#2}{}{(#2)}}}

\providecommand*{\newsmartprefix}[2]{\wgmWarn{\newsmartprefix}}
\providecommand*{\newsmartsprefix}[2]{\wgmWarn{\newsmartsprefix}}
\providecommand*{\newdefinition}[1]{\wgmWarn{\newdefinition}\newdefinitionaux}
\providecommand*{\newdefinitionaux}[2][]{}
\newcommand{\ands}{\fbox{this should not be used, is just to prevent 
    unobserved redefinition of command \string\ands}}
\newcommand*{\sequent}[4][]{\gdef\and{\quad\hfill}
    \gdef\ands{\;\hfill\ldots\hfill\;}\global\setlabel{#1}
    {#1}\frac{\:\:#2\:\:}{\;#3\;}\;\text{\small%
\begin{tabular}{@{}l@{}} #4 \end{tabular}}}
\providecommand*{\pair}[2]{{\langle #1, \, #2 \rangle}}
\providecommand*{\triple}[3]{{\langle #1, \, #2, \, #3 \rangle}}

\newcommand*{\R}{\mathbb{R}}
\newcommand*{\Conf}{\mathit{Conf}}
\newcommand*{\clauseif}{\ensuremath{\mathrel{\mathord{:}\mathord{-}}}}
\newcommand*{\ra}{\rightarrow}
\newcommand*{\nra}{\not\rightarrow}
\newcommand*{\tccp}{\textit{tccp}}
\newcommand*{\ccp}{\textit{ccp}}
\newcommand*{\cc}{\textit{cc}}
\newcommand*{\hcc}{\textit{hcc}}
\newcommand*{\hytccp}{{\textsc hy}-\-\tccp}
\newcommand*{\askip}{\mathord{\mathsf{stop}}}
\newcommand*{\atell}{\parensmathoper{\mathsf{tell}}}
\newcommand*{\aask}[1]{\mathop{\mathsf{ask}}\ifempty{#1}{}{(#1)\ra}}
\newcommand*{\asumask}[4][i]{\sum_{#1=1}^{#2}\aask{#3_{#1}}{#4}_{#1}}
\newcommand*{\anow}[3]{\tagnow\ifempty{#1}{} {\: #1 \tagthen #2\ifempty{#3}{}{\tagelse #3}}}

\newcommand*{\ahiding}[3][]{\mathop{\exists^{#1}{#2}} {#3}}

\newcommand*{\tagelse}{\mathrel{\mathsf{else}}}
\newcommand*{\tagnow}{\mathop{\mathsf{now}}}
\newcommand*{\tagthen}{\mathrel{\mathsf{then}}}
\newcommand*{\aaskc}[1]{\tagaskc \ifempty{#1}{}{(#1)}}
\newcommand*{\achange}[3]{\tagchange \ifempty{#1}{}{(#1,#2,#3)}}
\newcommand*{\tagchange}{\mathop{\mathsf{change}}}
\newcommand*{\tagaskc}{\mathop{\widetilde{\mathsf{ask}}}}
\newcommand*{\tagask}{\mathop{\mathsf{ask}}}
\newcommand*{\taghence}{\mathop{\mathsf{hence}}}

\newcommand*{\CSdom}{\mathcal{C}} 
\newcommand*{\CSfalse}{\mathit{false}}
\newcommand*{\CShid}[2]{\mathop{\exists}\nolimits_{#1} #2}
\newcommand*{\CSimp}{\mathrel{\vdash}}
\newcommand*{\CSnimp}{\mathrel{\nvdash}}

\newcommand*{\CSmerge}{\mathbin{\wedge}}
\newcommand*{\CSord}{\preceq}
\newcommand*{\CStrue}{\mathit{true}}
\newcommand*{\CSys}{\mathbf{C}}
\newcommand*{\Var}{\mathit{Var}}

\newcommand*{\lub}{\ensuremath{\mathit{lub}}}
\newcommand*{\dom}{\parensmathoper{dom}} 
\newcommand*{\ecrs}{\epsilon}
\newcommand*{\true}{\mathit{true}}
\newcommand*{\HVar}{\widetilde{\mathit{Var}}}
\newcommand*{\HVarTuple}[3]{#1\mapsto(#2,#3)}
\newcommand*{\HVarUpdate}[2]{\ifempty{#1}{\lhd}{#1 \lhd #2}}
\newcommand*{\inv}{\mathit{inv}}
\newcommand*{\dt}{\sigma}
\newcommand*{\rad}{\ra_{\dt}}
\newcommand*{\rac}[1]{\ra_{\mathit{#1}}}
\newcommand*{\rag}[1]{\ra_{\mathit{#1}}}

\newcommand*{\nrag}[1]{\nra_{\mathit{#1}}}
\newcommand*{\hst}[2]{\langle #1,\, #2 \rangle}
\newcommand*{\Confc}{\widetilde{\mathit{Conf}}}
\newcommand{\cooler}{\texttt{cooler}}
\newcommand{\init}{\texttt{init}}

\newcommand{\control}{\texttt{controller}}
\newcommand{\off}{\mathit{off}}
\newcommand{\on}{\mathit{on}}

\newcommand{\acat}{\texttt{cat}}
\newcommand{\amouse}{\texttt{mouse}}

\renewcommand*{\askip}{\textsf{stop}}
\renewcommand*{\CSmerge}{\wedge}
\newcommand*{\hstdom}{\Gamma}
\newcommand*{\HCSdom}{\tilde{\mathcal{C}}}
\newcommand*{\HCSfalse}{\widetilde{\CSfalse}}
\newcommand*{\HCStrue}{\widetilde{\CStrue}}
\newcommand*{\HCSimp}{\tilde{\CSimp}}
\newcommand*{\HCSnimp}{\tilde{\CSnimp}}
\newcommand*{\HCShid}[2]{\mathop{\tilde{\exists}}\nolimits_{#1} #2}

\newcommand*{\HCSmerge}{\mathbin{\tilde{\CSmerge}}}

\newcommand*{\loc}{\mathit{loc}}
\newcommand*{\lab}{\Lambda}
\newcommand*{\Loc}{\mathit{Loc}}
\newcommand*{\Init}{\mathit{Init}}
\newcommand*{\Inv}{\mathit{Inv}}
\newcommand*{\Flow}{\mathit{Flow}}
\newcommand*{\Jump}{\mathit{Jump}}

\begin{document}
    
\maketitle

\begin{abstract}
    Hybrid systems, which
    combine discrete and continuous dynamics, require quality modeling languages to be either described or analyzed.
    The Concurrent Constraint paradigm (\ccp) is an expressive declarative paradigm,
    characterized by the use of a common constraint store to 
    communicate and synchronize concurrent agents.
    In this paradigm, the information is stated in the form of constraints,
    in contrast to the variable/value style typical of imperative languages.
    Several extensions of \ccp{} have been proposed in order to model
    reactive systems. One of these extensions is the
    Timed Concurrent Constraint Language (\tccp)
    that adds to \ccp{} a notion of discrete time and
    new features to model time-out and preemption
    actions.
    
    The goal of this paper is to explore the expressive power of \tccp{}
    to describe hybrid systems.
    We introduce the language
    \hytccp{} as a conservative extension of \tccp, by adding a notion of continuous time
    and new constructs to describe the continuous dynamics of hybrid systems.
    %in the style of hybrid automata.
    In this paper, we present the syntax and the operational semantics of \hytccp{}
    together with some examples that show the expressive power of our new language.
\end{abstract}

\section{Introduction}
\label{sec:intro}
In the last years, concurrent, reactive and hybrid systems
have become essential
to model a large number of modern applications.
Often, systems of this kind are classified as critical,
\ie{} an error in the software can have
tragic consequences in terms of human lives or money.
This is the case of avionic or automotive software,
e-banking, or financial applications.

Description, verification and analysis of concurrent and reactive systems
are very hard tasks, due to the concurrent execution of different agents
and to issues of synchronization.
In the case of hybrid systems, these phases are even harder due to the combination of discrete 
and continuous dynamics and the presence of real-valued variables.
Therefore, it is important to develop
high-level description languages that allow
these systems to be modeled with
enough precision and at the same time
that ease the application of formal methods techniques.

Many formalisms have been developed to describe concurrent systems.
One of these is the \emph{Concurrent Constraint paradigm} (\ccp)~\cite{Saraswat89phd},
a simple but powerful model for concurrent systems.
It differs from other paradigms mainly due to the
notion of store-as-constraint that replaces the classical store-as-valuation model.
In this paradigm, the agents running in parallel
communicate by means of a global constraint store.
The \emph{Timed Concurrent Constraint Language}
\cite{deBoerGM99} (\tccp{} in short) is a concurrent logic language obtained
by extending \textit{ccp} with the notion of time and a suitable
mechanism to model time-\-outs and preemptions.

In this paper, we present the language \hytccp{}: an extension of \tccp{} over continuous time.
\hytccp{} is a non-deterministic and synchronous language that incorporates
continuous variables that follow dynamics determined by an ordinary differential equation (ODE).
Its declarative nature facilitates
a high level description of hybrid systems
in the style of hybrid automata \cite{Henzinger96}.
Furthermore, its logical nature
facilitates the development of semantics
based program manipulation tools
for hybrid systems
(verifiers, analyzers, debuggers\dots).
Parallel composition of hybrid automata is naturally
supported in \hytccp{} due to the existence of a global shared store and to
the synchronization mechanism inherited from \tccp.
By defining \hytccp{}, we show that the extension of a declarative constraint
language with continuous dynamics is not only possible,
but it leads to a powerful and expressive
language able to describe complex hybrid systems.

In this paper, we have only considered the modeling of multi-rated \cite{DawsY95} hybrid systems,
\ie{} systems whose continuous variables follow a constant dynamics.
However, in the future we aim to
relax this restriction in order to describe
more complex dynamics such as those defined by rectangular sets.

The paper is organized as follows.
In \smartref{sec:background}, we briefly introduce the language \tccp{}
and the essential aspects of hybrid automata. % \cite{Raskin05}.
In \smartref{sec:extension}, we introduce the new language \hytccp{} together with
its operational
semantics, and we describe 
the new features that have been added to \tccp{} in order to model hybrid systems.
\smartref{sec:examples} contains some examples to highlight the
expressive power of \hytccp.
\smartref{sec:rel_work} presents some related work and, finally,
\smartref{sec:conclusions} concludes the paper and outlines future work.

\section{Background}
\label{sec:background}
In this section we present some background to clarify the contributions of the paper.
In Subsection~\ref{subsec:tccp}, 
we introduce the language \tccp, the starting point for the definition of \hytccp.
In Subsection~\ref{subsec:hybrid}, we introduce the basic notions of hybrid automata, which is the 
formalism commonly used to describe hybrid systems.

\subsection{The Timed Concurrent Constraint Language}\label{subsec:tccp}

The \emph{Timed Concurrent Constraint Language} (\tccp, \cite{deBoerGM99})
is a time extension of \ccp{}.
It adds to \ccp{} the notion of time and the ability to capture the
absence of information. With these features, one can specify behaviors
typical of concurrent and reactive systems.

The computation in  \tccp{}
proceeds as the concurrent execution
of several agents that can monotonically
add constraints in
a global \emph{store}
or query information from it.
As are all the languages from the \cc{} paradigm,
\tccp{} is parametric
\wrt\ a \emph{cylindric
constraint system}.

\begin{definition}[Cylindric constraint system \cite{deBoerGM99}]\label{def:CylCS}
\index{cylindric constraint system}
    A cylindric constraint system is an algebraic structure of the
    form:
    $$\CSys= \langle \CSdom,\, \CSord,\, \CSmerge,\, \CStrue,\, \CSfalse,\, \Var,\, \exists \rangle $$
    such that:
    \begin{enumerate}
        \item $\langle \CSdom,\, \CSord,\, \CSmerge,\, \CStrue,\, \CSfalse \rangle$
        is a complete lattice where
        $\CSmerge$ is the least upper bound ($lub$)
        operator, and
        $\CStrue$ and $\CSfalse$ are, \resp, the least and the greatest
        elements of $\CSdom$. We often
        use the inverse order $\vdash$ (the \emph{entailment} relation) instead
        of $\CSord$ over constraints.  Formally $\forall c,d\ \in \CSdom$
        $c \CSord d\Leftrightarrow d \CSimp c$.
        \item $\Var$ is a
        denumerable set of variables.
        \item For each element $x\in \Var$, a
        function (also called cylindric operator) $\exists_{x} \colon
        \CSdom\ra \CSdom$ is defined such that, for any $c,d\in \CSdom$ the
        following axioms hold:
        \begin{enumerate}
            \item $c\CSimp \exists_{x}c$
            \item if $c\CSimp d$ then $\exists_{x}c\CSimp
            \exists_{x}d$
            \item $\exists_{x}(c\CSmerge
            \exists_{x}d)=\exists_{x}c\CSmerge
            \exists_{x}d$
            \item
            $\exists_{x}(\exists_{y}c)=\exists_{y}(\exists_{x}c)$
        \end{enumerate}
    \end{enumerate}
\end{definition}

The entailment relation $\CSimp$ intuitively states that 
if $c$ contains more information than $d$ then $c\CSimp d$.
The $\lub$ operator
$\CSmerge$ merges the information from two constraints
(\eg{} $x>0 \CSmerge x>5 \CSmerge y=9 := x>5 \CSmerge y=9$ and $x=0 \wedge x=7 := \CSfalse$).
The \emph{cylindrification} (or \emph{hiding}) operator is defined in terms of a
general notion of existential quantifier. It is used to project away information
about the considered variable in order to make it local
to the constraint and hide it from the context
(\eg{} $\CShid{x}{(x=0 \wedge y=x \wedge z>7)} := y=0 \wedge z>7$).

The \tccp{} global store is \emph{monotonic} in the sense that
once a constraint is added to the store, it cannot be removed.
Thus, given the store $x>0 \CSmerge y>2$ we can add the information $x>5$
and obtain the store
$x>5 \CSmerge y>2$.
Furthermore, by adding $x=0$ we obtain the inconsistent store $\CSfalse$
since the constraint $x=0$ is in contradiction with the
information already present in the store.

The syntax of \tccp{} agents is given by the grammar:
\begin{align*}
    A ::= \askip \mid \atell{c} \mid A \parallel A \mid
    \ahiding{x}{A} \mid
    \textstyle{\sum_{i=1}^{n}\aask{c_{i}} A}
    \mid \anow{c}{A}{A}
    \mid p(\vec{x})
\end{align*}
where $c$, $\seq{c}$ are finite constraints in $\CSdom$,
$p$ is a process symbol,
and $\vec{x} \in \Var\times\dots\times\Var$.
A \tccp{} program
is a pair $D\,.\,A$, where $A$ is the initial agent
and $D$ is a set of \emph{process declarations} of
the form $p(\vec{x}):- A$.

The \emph{operational semantics} of \tccp{} \cite{deBoerGM99}
is described by a transition system $T=(\Conf, \ra)$.
Configurations in $\Conf$ are pairs $\pair{A}{c}$ representing the agent
$A$ to be executed in the current global store $c$.  The transition
relation ${\ra} \subseteq \Conf\times\Conf$ is the least relation
satisfying the rules in Figure~\ref{fig:op_sem_tccp}.
Each transition step takes
exactly one time-unit.
The notion of time is introduced by defining a global clock which
synchronizes all agents.

As can be seen from the rules, the $\askip$ agent represents the successful
termination of the computation.
The $\atell{c}$ agent adds the constraint $c$ to the current store
by means of the $\CSmerge$ operator 
and then stops.
It takes one time-unit, thus the constraint $c$ is visible to the other
agents from the following time instant.
The choice agent $\asumask{n}{c}{A}$ consults the store and
non-deterministically executes
(at the following time instant)
one of the agents $A_i$ whose corresponding guard $c_i$ is entailed by the current
store; otherwise, if no guard is entailed by the store, the agent suspends.
The conditional agent $\anow{c}{A}{B}$ behaves (in the current time instant)
like $A$ (\resp\ $B$) if $c$ is (\resp\ is not) entailed by the store.
This conditional agent is able to process
\emph{negative information} (lack of some information): it can capture
when some information is not present in the store since the agent $B$
is executed both when $\neg c$ is satisfied, but also when neither $c$
nor $\neg c$ are satisfied.
$A\parallel B$ models the parallel composition of $A$ and $B$ in terms of
maximal parallelism,
\ie\ all the enabled agents of $A$ and $B$ are executed at the same time.
The agent $\ahiding{x}{A}$ makes variable $x$ local to $A$, to this end,
it uses the $\CShid{}{}$ operator of the constraint system.
More specifically, it behaves like $A$ with $x$ considered local, \ie\ the
information on $x$ provided by the external environment is hidden to $A$,
and the information on $x$ produced by $A$ is hidden to the external world.
In the corresponding rule, the store $l$ in the agent $\ahiding[l]{x}{A}$
represents the store local to $A$.  This auxiliary operator is linked to
the hiding construct by setting the initial local store to $\CStrue$, thus
$\ahiding{x}{A} \dfn \ahiding[\CStrue]{x}{A}$.
Finally, the agent $p(\vec{x})$
takes from $D$ a declaration of the form
$p(\vec{x}):-{A}$ and then executes $A$.

\begin{figure}[t]
    \begin{minipage}{\textwidth}
        {\normalsize
        {\setlength{\jot}{1ex}
        \begin{align*}
            & \sequent{}
            {\pair{\atell{c}}{d} \ra \pair{\askip}{c \CSmerge d}
            }
            {}{}
            & \sequent{\exists\,1 \leq j \leq n.\, d \CSimp c_{j} }
            {\pair{\asumask{n}{c}{A}}{d} \ra \pair{A_{j}}{d} }
            {}{}
            \\
            & \sequent{\pair{A}{d} \ra \pair{A'}{d'},\, d\CSimp c }
            {\pair{\anow{c}{A}{B}}{d} \ra \pair{A'}{d'} }
            {}{}
            & \sequent{ \pair{A}{d} \nra,\, d\CSimp c }
            { \pair{\anow{c}{A}{B}}{d} \ra
            \pair{A}{d} }
            {}{}
            \\
            & \sequent{ \pair{B}{d} \ra \pair{B'}{d'},\, d\CSnimp c }{
            \pair{\anow{c}{A}{B}}{d} \ra \pair{B'}{d'} }
            {}{}
            & \sequent{ \pair{B}{d} \nra,\, d\CSnimp c }{ \pair{\anow{c}{A}{B}}{d} \ra
            \pair{B}{d} }
            {}{}
            \\
            & \sequent{ \pair{A}{d} \ra \pair{A'}{d'} \quad \pair{B}{d} \ra
            \pair{B'}{c'} }{ \pair{A\parallel B}{d} \ra \pair{A'\parallel
            B'}{d'\CSmerge c'} }
            { }{}
            & \sequent{ \pair{A}{d} \ra \pair{A'}{d'} \quad \pair{B}{d} \nra }{
            \begin{array}{c}
                \pair{A\parallel B}{d} \ra \pair{A'\parallel B}{d'} \\
                \pair{B\parallel A}{d} \ra \pair{B \parallel A'}{d'}
            \end{array}
            }{ }{}
            \\
            & \sequent{ \pair{A}{l \CSmerge \CShid{x}{d}} \ra\pair{B}{l'} }{
            \pair{\ahiding[l]{x}{A}}{d} \ra \pair{\ahiding[l']{x}{B}}{d
            \CSmerge \CShid{x}{l'}} }{ }{}
            & \sequent{ p(\vec{x})\clauseif A \in D }{ \pair{ p(\vec{x}) }{d} \ra \pair{A}{d}
            }{}{}
        \end{align*}
        }
        }
        \caption[The transition system for \tccp{}.]{The transition system for \tccp{}.}
        \label{fig:op_sem_tccp}
    \end{minipage}
\end{figure}

\subsection{Introduction to hybrid automata} \label{subsec:hybrid}

Many real systems have complex behaviors and evolve following both
discrete and continuous dynamics. These systems are called hybrid systems.
For instance, a cooler system is a hybrid system: it has two discrete states
(\emph{on} or \emph{off}) that are chosen
according to the temperature of the room, which evolves continuously over time.

\emph{Hybrid automata} \cite{Henzinger96} are an extension of finite-state automata
used to describe hybrid systems.
Intuitively, the discrete behavior of a hybrid automaton is defined by means of a finite set of discrete states
(called {\em locations}) and a set of (instantaneous) \emph{discrete transitions} from one location to another.
The continuous behavior of hybrid automata
is described at each location by means of some
Ordinary Differential Equations (ODEs) which describe
how continuous variables evolve over time
(\emph{continuous transitions}).

\begin{definition}[Hybrid automaton]\label{def:hybrid automata}
A \emph{hybrid automaton} $H$ is a tuple
$$\langle \Loc,\, T,\, \Sigma,\, X,\, \Init,\, \Inv,\, \Flow, \Jump \rangle$$ where:

\begin{itemize}
    \item $\Loc$ is a finite set $\{\loc_1,\dots,\loc_n\}$ of discrete states
    (locations).
    
    \item $T \subseteq \Loc \times \Loc$ is a finite set of
    discrete transitions.
    
    \item $\Sigma$ is a set of event names, associated with a labelling function
    $\lab:T \to \Sigma$.
    
    \item $ X = \{x_1,\dots,x_m\}$ is a finite set of real-valued
    variables.
    The set $\dot{X} = \{\dot{x_1},\dots,\dot{x_m}\}$
    represents the first derivatives of the elements in $X$.
    In addition, the set $ X' = \{x'_1,\dots,x'_m\}$ represents the updates of
    the variables when a discrete transition takes place. In this section,
    we assume that discrete variables are continuous variables whose
    derivative is zero at all locations.
    
    \item The functions $Init$, $Inv$ and $Flow$ assign
    predicates to each location $\loc \in \Loc$. $\Init(\loc)$
    establishes the possible initial values for the continuous variables
    at location $\loc$.
    $\Inv(\loc)$ constrains the values of the
    continuous variables at location $\loc$.
    $\mathit{\Flow(\loc)}$ contains the differential
    equations describing the evolution of the
    continuous variables at location $\loc$.
    
    \item Function $\Jump$ assigns to each discrete transition $t \in T$
    a \emph{guard}
    that must be satisfied in order to allow the
    transition to take place,
    and a \emph{reset} predicate which
    updates the value and/or the flow of a continuous variables.
    \end{itemize}
\end{definition}

\begin{example}
   \smartref{fig:cooler} shows a hybrid automaton
   modeling a cooler system. The automaton has two locations $\on$ and $\off$
   and a continuous variable $T$ storing the room temperature.
   When the automaton is at location $\on$ (the cooler is turned on)
   the temperature decreases at rate $-0.5$. When the location is $\off$
   (the cooler is turned off) the temperature increases at rate $+2.0$.
   Transitions between locations represent the turning on or off of the cooler.
   These transitions are guarded with conditions. For instance, transition
   $\on$-$\off$ takes place when the temperature is $26$, while transition
   $\off$-$\on$ takes place when the temperature is $30$.
   
   \begin{figure}
       \begin{center}
           \begin{tikzpicture}[->,>=stealth',shorten >=1pt,auto,node distance=5cm,
                               semithick]
             \tikzstyle{every state}=[text=black,very thick]
           
             \node[state] (A) [label = above:$T\geq 26$, label = below:$\mathit{on}$] {$\dot{T} = -0.5$};
             \node[state] (B) [right of=A, label = above:$T\leq 30$, label = below:$\mathit{off}$] { $\dot{T} = +2.0$};
           
             \path (A) edge [bend left] node {$T = 26$} (B)
                   (B) edge [bend left] node {$T = 30$} (A);
           \end{tikzpicture}
       \end{center}
     \caption{Hybrid automaton for the cooler system}
     \label{fig:cooler}
   \end{figure}
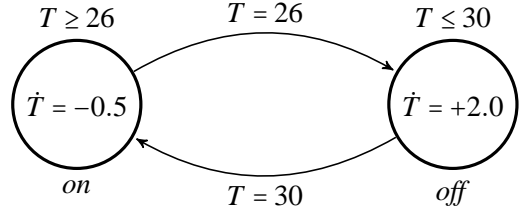
\end{example}

A hybrid automaton behaves like a \emph{timed transition system}
(TTS), where each step is labelled either with a positive real value $\tau$ (continuous transition
of duration $\tau$) or with $\dt$ (discrete transition).
Let $[X \rightarrow \mathbb{R}]$ be the set
of maps from $X$ to $\mathbb{R}$. An automaton state, called
hybrid state from now on, is a pair $(\loc,v) \in (Loc \times [X \to \mathbb{R}])$, where
$\loc\in \Loc$ is a location of the automaton, and $v \in [X \rightarrow\mathbb{R}]$
maps each continuous variable to its current value.

Let $p$ be a predicate over $X \cup \dot{X}$ or $X \cup X'$,
then $\lsem p \rsem$ denotes all functions $v \in [X \rightarrow \mathbb{R}]$
that satisfy $p$.

\begin{definition}[Trajectories]\label{def:trajectories}
Let $H=\langle \Loc,\, T,\, \Sigma,\, X,\, \Init,\, \Inv,\, \Flow, \Jump \rangle$ be a hybrid
automaton. We consider two types of transitions:
\begin{description}
    \item[Discrete transitions]  Let $(\loc,\loc') \in T$,
    ${(\loc,v) \rad (\loc',v')}$, iff $v,v' \in [X \rightarrow \mathbb{R}]$,
    and $(v,v') \in \lsem \Jump(t)\rsem$.
    \item [Continuous transitions] For each $\tau \in
    \R^+$, we have $(\loc,v) \rac{\tau} (\loc,v')$
    iff there exists a differentiable function $f:[0,\tau] \rightarrow
    \mathbb{R}^m$, $\dot{f}:[0,\tau] \rightarrow \mathbb{R}^m$ being
    its first derivative, such that:
        \begin{itemize}
            \item $f(0) = v$
            \item $f(\tau) = v'$
            \item $\forall \tau' \in [0,\tau], f(\tau') \in \lsem \Inv(\loc)\rsem$
            \item $(f(\tau'),\dot{f}(\tau')) \in \lsem\Flow(\loc)\rsem$
        \end{itemize}
\end{description}

A \emph{trajectory} is a (possible infinite) sequence of hybrid states
such as $(\loc_0,v_0) \rag{\lambda_1} (\loc_1,v_1) \rag{\lambda_2} \dots \rag{\lambda_n} (\loc_n,v_n) \rightarrow_{\lambda_n} \dots$,
where for all $i\geq0$, $v_i \in \lsem \Inv(\loc_i)\rsem$ and $\lambda_i \in \mathbb{R} \cup \{\dt\}$.
\end{definition}

It is worth noting that the system is free
to select non-deterministically at each moment
any enabled transition, either discrete or continuous.

\begin{example}\label{ex:thermoAut}
    Considering the hybrid system in \smartref{fig:cooler},
    the following trajectory represents a possible
    evolution of the automaton starting at hybrid state $(\on,27)$:
    $(\on,27)\rac{1} (\on,26.5) \rac{1} (\on,26) \rad (\off,26)\rac{0.5}(\off,27)\rac{1.5}(\off,30)\rad (\on,30) \dots$
\end{example}

\section{\hytccp{}: an extension of \tccp{} over continuous time}
\label{sec:extension}
In this section, we present the language \hytccp,
which subsumes \tccp{} and includes new agents in order to model
the continuous behavior typical of hybrid systems in the style of hybrid automata.
In contrast to \tccp{},
in \hytccp{} we consider a notion of \emph{continuous} time
by means of a global continuous clock.

\hytccp{} uses a \tccp{} monotonic store (called \emph{discrete store}) to model the
information about the current location and the associated
invariants of a hybrid automaton.
Discrete transitions are
modeled as instantaneous
transitions in \hytccp{} and they are used to synchronize parallel agents/automata.
In summary, the features offered by \tccp{} are used to model the discrete behavior
of hybrid automata.
However, hybrid automata are characterized by the use of continuous variables whose values
change following some ODEs.
For this reason, the \tccp{} store is extended by adding a
component called \emph{continuous store}.
The continuous store is not monotonic, instead it records the dynamical evolution of the continuous
variables.

We distinguish the set of discrete variables $\Var$, whose
information is accumulated monotonically,
and the set of continuous variables $\HVar$,
whose values change continuously over time ($\Var\cap\HVar = \emptyset$).
Constraints in $\CSdom$ are now defined over $\Var\cup\HVar$.

A \emph{continuous store}
is a function that associates
a continuous variable with two real numbers: its value and its flow,
which indicates how its value changes over time.
In this work, we consider only ODEs of the form $\dot{x} = n$
with $n\in\mathbb{R}$. In the future, we intend to also consider ODEs of the form
$\dot{x} \in [n_1,n_2]$ with $n_1,n_2\in\mathbb{R}$ in order to model rectangular hybrid systems.

We denote as $\HCSdom = [\HVar \hookrightarrow (\R\times\R)]$
the set of all possible continuous stores, and as
$\HCStrue$ and $\HCSfalse$ the empty and the inconsistent continuous store, \resp.
We denote with $\dom{\tilde{c}} \subseteq \HVar$ the domain of $\tilde{c}$.
Given $\tilde{c} \in \HCSdom$ and $x \in \dom{\tilde{c}}$,
$\tilde{c}(x) = \pair{v}{f}$ means that $x$ has value $v$ (denoted as $\tilde{c}(x).v$) and flow $f$
(denoted as $\tilde{c}(x).f$).
The binary operator $\HCSmerge: \HCSdom \times \HCSdom \ra \HCSdom$
merges the information from two continuous stores. In the case the same
variable appears in both stores with different values or flows, their
merge is inconsistent. Given $\tilde{c}, \tilde{d}\in \HCSdom$:
\begin{align*}
    &\tilde{c} \HCSmerge \HCStrue = \tilde{c} \quad \tilde{c} \HCSmerge \HCSfalse = \HCSfalse\\ % \quad
    &\tilde{c} \HCSmerge \tilde{d} = \HCSfalse \quad \text{if } \exists x \in \dom{\tilde{c}}\cap\dom{\tilde{d}}.\
    \tilde{c}(x) \neq \tilde{d}(x)\\
    &\tilde{c} \HCSmerge \tilde{d} =\lambda y.
    \begin{cases}
        \tilde{c}(y)\quad \text{if } y\in\dom{\tilde{c}}\\
        \tilde{d}(y)\quad \text{if } y\in\dom{\tilde{d}}
    \end{cases} \text{if } \forall x \in \dom{\tilde{c}}\cap\dom{\tilde{d}}.\ \tilde{c}(x) = \tilde{d}(x)
\end{align*}
We define the operator $\HCShid{}{}: \Var\times\HCSdom \ra \HCSdom$ such that,
given $\tilde{c} \in \HCSdom$
and $x\in\HVar$,
$\HCShid{x}{\tilde{c}}$ deletes the information about $x$ in $\tilde{c}$.

Given $\tilde{c} \in \HCSdom$,
$x \in dom(\tilde{c})$ and $v \in \R$, we
denote as $\tilde{c}[v/x]$ the continuous store that is equal
to $\tilde{c}$
except for the value of $x$ that becomes $v$.
\begin{align*}
    & \tilde{c}[v/x] =\lambda y.
    \begin{cases}
        \tilde{c}(y)\quad \text{if } y\in\dom{\tilde{c}}, y \neq x\\
        (v,\tilde{c}(x).f) \quad \text{if } y = x
    \end{cases}
\end{align*}

A \hytccp{} store is a pair $\hst{c}{\tilde{c}}$ where
$c\in\CSdom$ (discrete store) is a monotonic constraint store as in \tccp{}
and $\tilde{c}\in\HCSdom$ (continuous store)  is such that
$c \CSmerge \bigwedge_{x \in \dom{\tilde{c}}}
(x = \tilde{c}(x).v) \neq \CSfalse$, \ie{}
discrete and continuous store are consistent\footnote{We assume that our underlying constraint system
handles equality constraints.}.
We denote as $\hstdom$ the set of all possible \hytccp{} stores.
We define the extension of the entailment relation $\CSimp$
over \hytccp{} stores as $\HCSimp: \hstdom \times \CSdom$ such that
given $\hst{c}{\tilde{c}} \in \hstdom$ and $d \in \CSdom$,
$\hst{c}{\tilde{c}} \HCSimp d$ if $c \CSmerge \bigwedge_{x \in \dom{\tilde{c}}}
(x = \tilde{c}(x).v) \CSimp d$.
In other words, a store $\hst{c}{\tilde{c}}$
entails a constraint $d$ if the discrete store $c$
merged with the projection of the current values of the continuous variables
entails $d$ in the underlying constraint system.

Given $\tau\in\mathbb R^+$ we denote as $\hst{c}{\tilde{c}_\tau}$ the
continuous projection of the store $\hst{c}{\tilde{c}}$ at time $\tau$:
the values of the continuous variables are
updated at time $\tau$, while the flows are unchanged:
$\tilde{c}_\tau = \lambda y.\, \tilde{c}[n_y/y] \text{ where } y\in\dom{\tilde{c}} \text{ and }
n_y = \tilde{c}(y).v + (\tilde{c}(y).f * \tau)$.
For instance consider the store $\hst{x>10}{y\mapsto (2,5)}$,
its projection at time $3$ is the store $\hst{x>10}{y\mapsto (17,5)}$.
We say that $\hst{c}{\tilde{c}_\tau}$
is a continuous projection of $\hst{c}{\tilde{c}}$ at time $\tau$ that satisfies $d$
(denoted as $\hst{c}{\tilde{c}} \rightsquigarrow_\tau^{d} \hst{c}{\tilde{c}_\tau}$)
if for all $\tau'\in[0,\tau]$ $\hst{c}{\tilde{c}_{\tau'}} \HCSimp d$.
For instance, the above projection satisfies $y>0$:
$\hst{x>10}{y\mapsto (2,5)} \rightsquigarrow_{3}^{y>0} \hst{x>10}{y\mapsto (17,5)}$.

The update operator $\HVarUpdate{}{}$, given $\tilde{c}, \tilde{d} \in \HCSdom$ such that
$\dom{\tilde{c}} \cap \dom{\tilde{d}}= \{x_1,\dots,x_n\}$,
updates $\tilde{c}$ with the information of $\tilde{d}$ as follows:
$\HVarUpdate{\tilde{c}}{\tilde{d}} := (\HCShid{x_1,\dots,x_n}{\tilde{c}}) \HCSmerge {\tilde{d}}$.
Note that it is impossible to obtain an inconsistent continuous store
since the common variables are hidden from $\tilde{c}$
and replaced by the new values and flows from $\tilde{d}$.

In order to model the typical behaviors of hybrid
systems we introduce two new constructs \wrt{} the syntax of \tccp{}: $\tagchange$ and $\tagaskc$.

The agent $\tagchange$ updates the current
continuous store with a new value and/or flow for a given continuous variable.
It roughly corresponds to the \emph{reset} predicate of hybrid automata.

Continuous transitions are modeled by the new construct $\aaskc{\inv}$
that makes continuous variables evolve over \emph{continuous} time while the invariant $\inv$ is satisfied.
The \tccp{} choice agent is extended by allowing the non-deterministic
choice between discrete and continuous transitions in the following way:

$$\textstyle{\sum_{i=1}^{n}\aask{c_{i}} A + \sum_{j=1}^{m} \aaskc{\inv_{j}}}.$$

Here, the $\tagaskc$ branches can be non-\-deterministically selected
when the corresponding invariant $inv_j$ is entailed in the current store.
The continuous variables evolve over time while $inv_j$ holds and
until another ask branch is selected.

The syntax of \hytccp{} agents is given by the following grammar:
\begin{align*}
    A ::=\ &\askip \mid \atell{c} \mid A \parallel A \mid
    \anow{c}{A}{A} \mid
    \ahiding{x}{A} \mid p(\vec{x}) \mid \\
    &\achange{y}{v}{f} \mid
    \textstyle{\textstyle{\sum_{i=1}^{n}\aask{c_{i}} A + \sum_{j=1}^{m} \aaskc{\inv_{j}}}}
\end{align*}
where $c$,
$c_{i}$ and $\inv_j$ are finite constraints in $\CSdom$, $y$ is a continuous variable in $\HVar$,
$v,\, f \in\R$,
$p$ is a process symbol, $x \in \Var\cup\HVar$,
$\vec{x} \in (\Var\cup\HVar) \times \dots \times (\Var\cup\HVar)$,
$n\geq 0$ and $m\geq 0$.

The \emph{operational semantics} of \hytccp{}
is described by a transition system $T=(\Confc, \rad, \rac{\tau})$.
Configurations in $\Confc$ are triples $\triple{A}{c}{\tilde{c}}$
representing the agent
$A$ to be executed in the current extended store $\hst{c}{\tilde{c}}$.
In contrast to the \tccp{} approach,
the discrete transition relation $\rad \subseteq \Confc\times\Confc$
does not represent the passage of one time unit.
Instead, it models a computational step which does not consume time
but it is needed to synchronize the agents in parallel.
The continuous passage of time is modeled by the
transition relation $\rac{\tau} \subseteq \Confc\times\Confc$ where
$\tau \in \R^+$ is a (strictly) positive real number
that indicates the duration of the transition.
In Figure~\ref{fig:op_sem_hytccp}, we formally
describe the operational semantics of \hytccp{}.
Wherever possible we will use the subindex $\lambda\in \R^+ \cup \{\dt\}$ to represent both kinds of transitions
(discrete and continuous).

Rule~\ref{rule:R1} shows the effects of adding a constraint $c\in\CSdom$ to
the current discrete store.
In Rule~\ref{rule:R1'}, the agent $\tagchange$ updates the continuous store $\tilde{d}$ with
a new initial value $v$ and a new flow $f$ for the variable $y$ by using the update operator $\HVarUpdate{}{}$.

Rules~\ref{rule:R2} and \ref{rule:R2'} describe the non-\-deterministic choice behavior.
Rule~\ref{rule:R2} represents the discrete transition that is performed when
one of the $\tagask$ guards is entailed in the current store.
In this case the corresponding agent is executed in the next step.
Rule~\ref{rule:R2'} models the continuous evolution of the system
while one of the $\tagaskc$ invariants holds in the store.
After a continuous transition of duration $\tau$,
the values of the variables in the continuous store $\tilde{d}$
are updated while the discrete store is unchanged.
At the end of that transition the non-\-deterministic choice is
executed again allowing another
discrete or continuous branch to be selected.
In the case no guard or invariant holds
this agent suspends.

Rules~\ref{rule:R3}, \ref{rule:R3'}, \ref{rule:R4} and \ref{rule:R4'}
describe the behavior of  agent $\anow{}{}{}$. This agent behaves as
$A$ if $c$ is entailed by the constraint store,
otherwise it behaves as $B$.

Rule~\ref{rule:R5} represents the parallel execution of
two discrete transitions in terms of maximal parallelism,
\ie{} all the enabled agents of $A$ and $B$ are executed at the same time.
Rule~\ref{rule:R6} represents the parallel execution of
two continuous transitions, note that their duration must coincide.
Rule~\ref{rule:R7} expresses the parallel composition of a discrete and a continuous transition.
In this case, the discrete transition is executed before the continuous one.
Rule~\ref{rule:R8} states that when an agent is blocked, the other one performs its transition
(discrete or continuous).

In Rule~\ref{rule:R9}, the agent $\ahiding[\hst{l}{\tilde{l}}]{x}{A}$ makes variable $x$ local to $A$.
It behaves like $A$ with $x$ considered local, \ie\ the
information on $x$ provided by the external environment is hidden from $A$ by using the $\HCShid{}{}$ operator,
and, in the same way, the information on $x$ produced by $A$ is hidden from the global environment.
The store $\hst{l}{\tilde{l}}$ in the agent $\ahiding[\hst{l}{\tilde{l}}]{x}{A}$
represents the store local to $A$.  This auxiliary operator is linked to
the hiding construct by setting the initial local store to $\hst{\CStrue}{\HCStrue}$, thus
$\ahiding{x}{A} \dfn \ahiding[\hst{\CStrue}{\HCStrue}]{x}{A}$.

Finally, in Rule~\ref{rule:R10}, the agent $p(\vec{x})$
takes from $D$ a declaration of the form
$p(\vec{x}) \clauseif A$ and executes $A$.

\begin{figure}[tp]
    \begin{minipage}{\textwidth}
        {\normalsize
        {\setlength{\jot}{1ex}
        \begin{align*}
            &\sequent{}
                {
                \triple{\atell{c}}{d}{\tilde{d}}
                \rad
                \triple{\askip}{c \CSmerge d}{\tilde{d}}
                }{}{\tag{\textbf{R1}}}\label{rule:R1}
            \\
            &\sequent{}
                {
                \triple{\achange{y}{v}{f}}{d}{\tilde{d}}
                \rad
                \triple{\askip}{d}{\HVarUpdate{\tilde{d}}{\HVarTuple{(y}{v}{f})}}
                }{}{\tag{\textbf{R1'}}}\label{rule:R1'}
            \\
            & \sequent{\exists\,1 \leq k \leq n\,.\, \hst{d}{\tilde{d}} \HCSimp c_{k} }
            {\triple{\textstyle{\sum_{i=1}^{n}\aask{c_{i}} A_i + \sum_{j=1}^{m}\aaskc{\inv_j}}}{d}{\tilde{d}}
             \rad
             \triple{A_k}{d}{\tilde{d}}
             }
            {}{\tag{\textbf{R2}}}\label{rule:R2}
            \\
            & \sequent{\exists\,1 \leq k \leq m,\,\tau\in\R^+ . \hst{d}{\tilde{d}} \rightsquigarrow_\tau^{inv_k}\hst{d}{\tilde{d}_{\tau}}}
            {\triple{\textstyle{\sum_{i=1}^{n}\aask{c_{i}} A_i + \sum_{j=1}^{m}\aaskc{\inv_j}}}{d}{\tilde{d}}
             \rac{\tau}
             \triple{\textstyle{\sum_{i=1}^{n}\aask{c_{i}} A_i + \sum_{j=1}^{m}\aaskc{\inv_j}}}{d}{\tilde{d}_\tau}
             }
            {}{\tag{\textbf{R2'}}}\label{rule:R2'}
            \\
            % now
            & \sequent{\triple{A}{d}{\tilde{d}}
             \rag{\lambda}
             \triple{A'}{d'}{\tilde{d}'} \quad \lambda \in \R^+ \cup \{\dt\} \quad \hst{d}{\tilde{d}} \HCSimp c}
            {\triple{\anow{c}{A}{B}}{d}{\tilde{d}}
             \rag{\lambda} \triple{A'}{d'}{\tilde{d}'}
             }
            {}{\tag{\textbf{R3}}}\label{rule:R3}
            \\
            & \sequent{\triple{A}{d}{\tilde{d}}
             \nrag{\lambda} \quad \lambda \in \R^+ \cup \{\dt\} \quad \hst{d}{\tilde{d}} \HCSimp c}
            {\triple{\anow{c}{A}{B}}{d}{\tilde{d}}
             \rad \triple{A}{d}{\tilde{d}}
             }
            {}{\tag{\textbf{R3'}}}\label{rule:R3'}
            \\
            & \sequent{\triple{B}{d}{\tilde{d}}
             \rag{\lambda}
             \triple{B'}{d'}{\tilde{d}'} \quad \lambda \in \R \cup \{\dt\} \quad \hst{d}{\tilde{d}} \HCSnimp c}
            {\triple{\anow{c}{A}{B}}{d}{\tilde{d}}
             \rag{\lambda} \triple{B'}{d'}{\tilde{d}'}
             }
            {}{\tag{\textbf{R4}}}\label{rule:R4}
            \\
            & \sequent{\triple{B}{d}{\tilde{d}}
             \nrag{\lambda} \quad \lambda \in \R \cup \{\dt\} \quad \hst{d}{\tilde{d}} \HCSnimp c}
            {\triple{\anow{c}{A}{B}}{d}{\tilde{d}}
             \rad \triple{B}{d}{\tilde{d}}
             }
            {}{\tag{\textbf{R4'}}}\label{rule:R4'}
            \\
             & \sequent{ \triple{A}{d}{\tilde{d}} \rad \triple{A'}{d'}{\tilde{d}'} \quad
             \triple{B}{d}{\tilde{d}} \rad \triple{B'}{d''}{\tilde{d}''} }{
                 \triple{A \parallel B}{d}{\tilde{d}} \rad \triple{A' \parallel B'}{d'\CSmerge d''}{\tilde{d}' \HCSmerge \tilde{d}''}
             }{}{\tag{\textbf{R5}}}\label{rule:R5}
            \\
            & \sequent{ \triple{A}{d}{\tilde{d}} \rac{\tau} \triple{A}{d}{\tilde{d}'} \quad
            \triple{B}{d}{\tilde{d}} \rac{\tau} \triple{B}{d}{\tilde{d}'} \quad \tau\in\R^+}{
                \triple{A \parallel B}{d}{\tilde{d}} \rac{\tau} \triple{A \parallel B}{d}{\tilde{d}'}
            }{}{\tag{\textbf{R6}}}\label{rule:R6}
           \\
           & \sequent{ \triple{A}{d}{\tilde{d}} \rad \triple{A'}{d'}{\tilde{d}'} \quad
           \triple{B}{d}{\tilde{d}} \rac{\tau} \triple{B}{d}{\tilde{d}''} \quad \tau\in\R^+}{
               \triple{A \parallel B}{d}{\tilde{d}} \rad \triple{A'\parallel B}{d'}{\tilde{d}'} 
           }{}{\tag{\textbf{R7}}}\label{rule:R7}
           \\
           & \sequent{ \triple{A}{d}{\tilde{d}} \rag{\lambda} \triple{A'}{d'}{\tilde{d}'} \quad
           \triple{B}{d}{\tilde{d}} \nrag{\lambda'} \quad \lambda, \lambda' \in \R^+ \cup \{\dt\} }{
           \begin{array}{c}
               \triple{A \parallel B}{d}{\tilde{d}} \rag{\lambda} \triple{A'\parallel B}{d'}{\tilde{d}'} \\
               \triple{B \parallel A}{d}{\tilde{d}} \rag{\lambda} \triple{B \parallel A'}{d'}{\tilde{d}'}
           \end{array}
           }{}{\tag{\textbf{R8}}}\label{rule:R8}
           \\
           & \sequent{ \triple{A}{l \CSmerge \CShid{x}{d}}{\tilde{l} \HCSmerge \HCShid{x}{\tilde{d}}}{}
           \rag{\lambda} \triple{B}{l'}{\tilde{l}'} \quad \lambda\in\R^+ \cup\{\dt\}}{
           \triple{\ahiding[\hst{l}{\tilde{l}}]{x}{A}}{d}{\tilde{d}}
           \rag{\lambda} \triple{\ahiding[\hst{l'}{\tilde{l}'}]{x}{B}}{d \CSmerge \CShid{x}{l'}}{\tilde{d} \HCSmerge \HCShid{x}{\tilde{l}'}} }{
           }\tag{\textbf{R9}}\label{rule:R9}
           \\
           & \sequent{ p(\vec{x}) \clauseif A \in D }{ \triple{ p(\vec{x}) }{d}{\tilde{d}} \rad \triple{A}{d}{\tilde{d}}
           }{}
           \tag{\textbf{R10}}\label{rule:R10}
           \end{align*}
          }
        }
        \caption[A fragment of the transition system for \hytccp{}.]{The transition system for \hytccp{}.}
        \label{fig:op_sem_hytccp}
    \end{minipage}
\end{figure}

Let us formalize the notion of behavior of a \hytccp{} program $P$
in terms of the transition system described in
\smartref{fig:op_sem_hytccp}. The small-step operational behavior of \hytccp{}
collects all the small-step computations
associated with $P$
(in terms of sequences of \hytccp{} stores closed by prefix)
for each possible
initial store.
We assume that subsequent continuous transitions are considered as a unique (maximal) one
whose length is equal to the sum of all the subsequent transition lengths.
For instance, a sequence of continuous transitions of the form
$\triple{A_0}{c_0}{\tilde{c}_0} \rac{\tau_1} \dots \rac{\tau_n} \triple{A_n}{c_n}{\tilde{c}_n}$
is considered as the unique transition
$\triple{A_0}{c_0}{\tilde{c}_0} \rac{\tau} \triple{A_n}{c_n}{\tilde{c}_n}$ where
$\tau = \sum_{i=1}^n \tau_i$\footnote{We assume that our system does
not exhibit Zeno behaviors.}.

\begin{definition}
    \label{def:ssBeha}
    Let $P=D\,.\,A$ be a \hytccp{} program.
    The \emph{small-step (observable) behavior} of $P$ is defined as:
    \begin{align*}
        \mathcal{B}^{\mathit{ss}}\lsem D\,.\,A \rsem&\dfn \bigcup_{\hst{c_0}{\tilde{c}_0}\in\hstdom} \big\{
        \hst{c_0}{\tilde{c}_0} \cdot  \hst{c_1}{\tilde{c}_1}
        \cdot \ldots \cdot \hst{c_n}{\tilde{c}_n}
        \mid \triple{A}{c_0}{\tilde{c}_0} \rag{\lambda_1}
        \triple{A_1}{c_1}{\tilde{c}_1}\\
        &\qquad \rag{\lambda_2}\dots \rag{\lambda_n}
        \triple{A_n}{c_n}{\tilde{c}_n},\, \forall 1\leq i \leq n.\ \lambda_n \in \R^+ \cup \{\dt\}
        \big\} \cup \{\ecrs\}
    \end{align*}
\end{definition}

\section{Examples}
\label{sec:examples}
In order to show the expressivity of \hytccp,
we present some examples of hybrid systems described in this language.
For each case, we present the \hytccp{} code and the corresponding hybrid automaton.

\subsection{Cooler system}

In \smartref{fig:coolingCode} we model in \hytccp{} the cooler system
introduced in \smartref{ex:thermoAut}.
The initial state of the cooler is set to $\off$ and the temperature $T$ initially has value $29$
and changes with a rate of $+2.0$.
The temperature value increases continuously over time (first $\aaskc{}$)
until the temperature is lower than or equal to the value of $30$.
When the temperature reaches this limit, the cooler is turned on and
the flow of the temperature changes from $+2.0$ to $-0.5$ (first $\aask{}$).
At this point, the temperature starts decreasing until it reaches the
value of $26$ (second $\aaskc{}$).
When this happens, the cooler is turned off and
the flow of the temperature is changed again to $+2.0$ (second $\aask{}$).

It is worth noting that, due to the monotonicity of the discrete constraint store, streams (written in a
list-fashion way) are used to model \emph{imperative-style} variables \cite{deBoerGM99}.
A stream is a list of the form
$\mathit{St} = [\on \mid \mathit{St}']$ where the head $\on$ represents the current value of $\mathit{St}$,
and the tail $\mathit{St}'$ is a free variable that will be instantiated with the future values of $\mathit{St}$.
Observe that we use the global constraint $T\geq 26 \wedge T\leq 30$ 
to add a global invariant of the cooler system ensuring that
the temperature always stays in the interval $[26,30]$.

\begin{figure}
{\scriptsize
\begin{center}
\begin{align*}
&\init \clauseif \exists\ St,T\ \big( \atell{St=[\off\mid\_]} \parallel \achange{T}{29}{+2.0}
\parallel \atell{ T\geq 26 \wedge T\leq 30 }  \parallel \cooler(St,T)
\big)\\
&\cooler(St,T) \clauseif \exists\ St' \Big(
\begin{aligned}[t]
    &\aaskc{St =[\off\mid\_] \wedge T \leq 30} \\
    +&\aask{St =[\off\mid\_] \wedge T = 30}
    \big(
    \atell{St =[\off\mid St']} \parallel \atell{St'=[\on\mid\_]} \parallel
    \achange{T}{30}{-0.5} \parallel \cooler(St',T)\big)\\
    + &\aaskc{St =[\on\mid\_] \wedge T\geq 26}\\
    +&\aask{St =[\on|\_] \wedge T=26}
    \big(
    \atell{St =[\on\mid St']} \parallel \atell{St'=[\off\mid\_]} \parallel
    \achange{T}{26}{+2.0} \parallel \cooler(St',T)\big)\Big)
\end{aligned}    
\end{align*}
\end{center}
\caption{\hytccp\ model for the cooler system}
\label{fig:coolingCode}
}
\end{figure}

The following partial trace represents the small-step behavior (see \smartref{def:ssBeha})
of $\cooler(St,T)$ starting from the initial store
$\hst{\mathit{St}=[\off\mid\_] \wedge T\geq26 \wedge T\leq30}{T\mapsto (29,+2.0)}$.
This means that, initially, the cooler is turned off and
the temperature has a value of $29$ and a flow of $+2.0$.
Moreover, the temperature is constrained to be between the
values $26$ and $30$.
Observe how the values $\on$ and $\off$ are accumulated in the stream $\mathit{St}$
in order to model the evolution of the state. The current state corresponds to
the last value added to the stream.
We use $\_$ to indicate that the tail of
the stream $\mathit{St}$ is a free variable
that can be instantiated with future values.
The continuous variables evolve over time until another discrete transition is executed.
The repeated equal stores occurring in the trace
correspond to the discrete computational steps taken in \hytccp{} (as well as in \tccp)
to evaluate one of the $\aask{}$ guards or to 
perform a procedure call. These steps are necessary to synchronize parallel
agents.
For sake of clarity, we explicitly indicate the kind of transition
occurring between two states (we write $\sigma$ for discrete transitions and the duration $\tau\in\R^+$ for
continuous ones).

{\footnotesize
\begin{align*}
    &\hst{\mathit{St}=[\off\mid\_] \wedge T\geq26 \wedge T\leq30}{T\mapsto (29,+2.0)} \cdot_{0.5}
    \hst{\mathit{St}=[\off\mid\_] \wedge T\geq26 \wedge T\leq30}{T\mapsto (30,+2.0)}\cdot_{\sigma}\\
    &\hst{\mathit{St}=[\off\mid\_] \wedge T\geq26 \wedge T\leq30}{T\mapsto (30,+2.0)}\cdot_{\sigma}
    \hst{\mathit{St}=[\off,\on\mid\_] \wedge T\geq26 \wedge T\leq30}{T\mapsto (30,-0.5)}\cdot_{\sigma}\\
    &\hst{\mathit{St}=[\off,\on\mid\_] \wedge T\geq26 \wedge T\leq30}{T\mapsto (30,-0.5)}\cdot_{8}
    \hst{\mathit{St}=[\off,\on\mid\_] \wedge T\geq26 \wedge T\leq30}{T\mapsto (26,-0.5)}\cdot_{\sigma}\\
    &\hst{\mathit{St}=[\off,\on\mid\_] \wedge T\geq26 \wedge T\leq30}{T\mapsto (26,-0.5)}\cdot_{\sigma}
    \hst{\mathit{St}=[\off,\on,\off\mid\_] \wedge T\geq26 \wedge T\leq30}{T\mapsto (26,+2.0)} \dots
\end{align*}}

\subsection{Cat and mouse race}

We consider the cat and mouse problem proposed in \cite{GuptaJSB94}
(see Figure~\ref{fig:catmouse} for the corresponding hybrid automaton).
The \hytccp{} code of this model is shown in Figure~\ref{fig:mouseCode}.
The positions of the cat and the mouse are modeled by two continuous variables,
called $C$ and $M$ \resp.
A mouse starts running from the point of origin
at a speed of 10 meters/second ($\achange{M}{0}{10.0}$)
towards a hole that is 100 meters away. After it has run 50 meters
it sends a signal to the cat ($\atell{\mathit{go}}$) and continues its run.
When the cat receives the signal $\mathit{go}$, it starts chasing the mouse
from the point of origin at a speed of 20 meters/second ($\achange{C}{0}{20.0}$).
The cat wins if it catches the mouse before it reaches the hole, otherwise it loses.
At the end of their run, the mouse and the cat send a message to the $\control$
($\mathit{end_M}$ and $\mathit{end_C}$, \resp), which decides
non-deterministically the winner
and informs of it through a signal ($\mathit{win_m}$ or $\mathit{win_c}$).
The winner, at this point, can claim his prize.

\begin{figure}
    \begin{center}
        \tiny{
        \begin{tikzpicture}[->,>=stealth',shorten >=1pt,auto, node distance=2.5cm, state/.style={circle,draw,minimum size=20pt},semithick]
            \tikzstyle{every state}=[text=black,very thick]
            \node[state] (E) [label = above:$\true$, label = below:$\mathit{loser}$] {$\dot{C} = 0.0$};
            \node[state] (C) [below left of=E, label = above:$\true$, label = below:$\mathit{finished}$] { $\dot{C} = 0.0$};
            \node[state] (B) [left of=C, label = above:$C \leq100$, label = below:$\mathit{chasing}$] { $\dot{C} = 20.0$}; 
            \node[state] (A)  [left of=B, label = above:$\true$, label = below:$\mathit{sleeping}$] {$\dot{C} = 0.0$};
            \node[state] (D) [below right of=C, label = above:$\true$, label = below:$\mathit{winner}$] {$\dot{C} = 0.0$};
            \path (A) edge node {$\mathit{go}$} (B)
                  (B) edge node {$C = 100 \mid \mathit{end_C}$} (C)
                  (C) edge node {$win_C$} (E)
                  (C) edge node {$win_M$} (D);
        \end{tikzpicture}}
        \tiny{
        \begin{tikzpicture}[->,>=stealth',shorten >=1pt,auto, node distance=2.5cm, state/.style={circle,draw,minimum size=20pt},semithick]
            \tikzstyle{every state}=[text=black,very thick]
            \node[state] (Am) [label = above:$M \leq50$, label = below:$\mathit{1st\_half}$] {$\dot{M} = 10.0$};
            \node[state] (Bm) [right of=Am, label = above:$M \leq100$, label = below:$\mathit{2nd\_half}$] { $\dot{M} = 10.0$};  
            \node[state] (Cm) [right of=Bm, label = above:$\true$, label = below:$\mathit{finished}$] { $\dot{M} = 0.0$};
            \node[state] (Dm) [above right of=Cm, label = above:$\true$, label = below:$\mathit{winner}$] {$\dot{M} = 0.0$};
            \node[state] (Em) [below right of=Cm, label = above:$\true$, label = below:$\mathit{loser}$] {$\dot{M} = 0.0$};  
            \path (Am) edge node {$M = 50 \mid \mathit{go}$} (Bm)
                  (Bm) edge node {$M = 100 \mid \mathit{end_M}$} (Cm)
                  (Cm) edge node {$win_M$} (Dm)
                  (Cm) edge node {$win_C$} (Em);
        \end{tikzpicture}}
    \end{center}
    \caption{Hybrid automata for the cat and mouse problem}
    \label{fig:catmouse}
\end{figure}
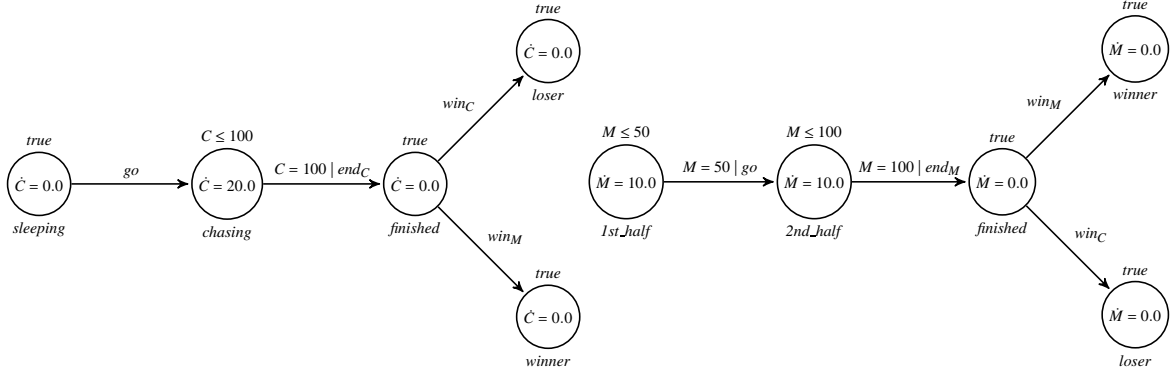

\begin{figure}
{\scriptsize
\begin{center}
\begin{align*}
    &\init \clauseif \amouse \parallel \acat \parallel \control\\
    % mouse
    &\amouse \clauseif \exists M\,\Big( \achange{M}{0}{10.0} \parallel \\
    &\qquad \begin{aligned}[t]
        &\big(\aaskc{M \leq 50}\\
        &+\aask{M=50} (\atell{\mathit{go}} \parallel
        \begin{aligned}[t]
        &(\aaskc{M\leq 100}\\
        &+\aask{M=100} (\atell{\mathit{end_m}} \parallel
        \begin{aligned}[t]
            &\aask{\mathit{win_m}}  \mathit{claimPrize(...)}\\
            &+\aask{\mathit{win_c}} \askip{}))) \big) \Big)
        \end{aligned}
        \end{aligned}
    \end{aligned}\\
    % cat
    &\acat \clauseif \exists C\, \Big(
    \aask{\mathit{go}}
    \begin{aligned}[t]
    &\big(\achange{C}{0}{20.0} \parallel\\
    &(
    \begin{aligned}[t]
        &\aaskc{C \leq 100}\\
        &+\aask{C=100}
        \begin{aligned}[t]
            (\atell{\mathit{end_c}} \parallel\,
            &\aask{\mathit{win_c}} \mathit{claimPrize(...)}\\
            &+\aask{\mathit{win_m}} \askip{}))\big)\Big)
        \end{aligned}
    \end{aligned}
    \end{aligned}\\
    % controller
    &\control \clauseif
    \begin{aligned}[t]
        &\aask{\mathit{end_m}} \atell{\mathit{win_m}}
        +
        \aask{\mathit{end_c}}  \atell{\mathit{win_c}}
    \end{aligned}
\end{align*}
\end{center}
\caption{\hytccp\  model for the cat and mouse race}\label{fig:mouseCode}
}
\end{figure}

\subsection{Gear shift system}

The hybrid automaton in \smartref{fig:gearAutomaton} represents a car gear shift system.
Each location models a gear ($1$, $2$ or $3$) and the fact that the speed is either increasing or decreasing
($\uparrow$ or $\downarrow$ \resp).
When the speed increases (\resp{} decreases) over time and it reaches a given
threshold, the current gear is changed to the upper (\resp{} lower) one.
When a signal of danger ($\mathit{dng}$)
is received, the system changes the current gear to the lower one and the speed starts decreasing.
At this point, when a signal of safe situation ($\mathit{safe}$) is received, the system is allowed to
stay in the current location as well as to
increase the speed. The latter case is modeled by the transitions from location $1\downarrow$ to 
location $1\uparrow$,
and from $2\downarrow$ to $2\uparrow$.

The \hytccp{} program modeling this system is shown in \smartref{fig:gearHytccp}.
The stream $G$ stores the evolution of the gear state.
The $\aaskc{}$ statements model the five locations of the automaton of 
\smartref{fig:gearAutomaton}, \ie{} the possible cases in which
a continuous transition is performed. It is worth noting that
the invariant of each location is modeled by the guard of the
corresponding
$\aaskc{}$ statement.
The first three $\aask{}$ statements model the 
\texttt{gearbox} shifting automatically into a higher (\resp{} lower) gear if
the speed $\mathit{V}$ reaches 
the upper (\resp{} lower) threshold of the current gear.
The \texttt{watcher} informs to the \texttt{gearbox} about the current external
situation (danger or safe), through channel $\mathit{WG}$.
When \texttt{gearbox} receives a danger signal $\mathit{dng}$
and the speed is growing
(fourth and fifth $\aask{}$ branches),
it moves to a lower gear, and changes the speed flow from positive to negative 
by means of a $\achange{}{}{}$ agent.
Otherwise, when it receives a safety signal $\mathit{safe}$
and the speed is decreasing
(sixth and seventh $\aask{}$ branches),
it is allowed to change the speed flow from negative to positive.

\begin{figure}
    \begin{center}
        \begin{tikzpicture}[->,>=stealth',shorten >=1pt,auto,node distance=5cm, semithick]
            \tikzstyle{every state}=[text=black,very thick]
          
            \node[state] (A) [label = above:$\mathit{V} \leq20$, label = below:$\mathit{1\uparrow}$] {$\dot{V} = +4.0$};
            \node[state] (B) [right of=A, label = above:$V \leq60 \wedge \neg\mathit{dng}$, label = below:$\mathit{2\uparrow}$] { $\dot{V} = +5.0$};
            \node[state] (C) [right of=B, label = above:$V \leq100 \wedge \neg\mathit{dng}$, label = below:$\mathit{3\uparrow}$] { $\dot{V} = +6.0$};
            \node[state] (D) [below of=A, label = above:$V\geq 0$, label = below:$\mathit{1\downarrow}$] {$\dot{V} = -4.0$};
            \node[state] (E) [below of=B, label = above:$V\geq 20$, label = below:$\mathit{2\downarrow}$] {$\dot{V} = -5.0$};
            \path (A) edge node {$V = 20$} (B)
                  (B) edge node {$V = 60$} (C)
                  (D) edge [bend right] node {$V\leq 20 \wedge \mathit{safe}$} (A)
                  (B) edge node {$\mathit{dng}$} (D)
                  (C) edge node {$\mathit{dng}$} (E)
                  (E) edge [bend right] node {$V\leq 60 \wedge \mathit{safe}$} (B)
                  (E) edge node {$V = 20$} (D);
        \end{tikzpicture}
    \end{center}
    \caption{Hybrid automaton for the gear shift system}
    \label{fig:gearAutomaton}
\end{figure}
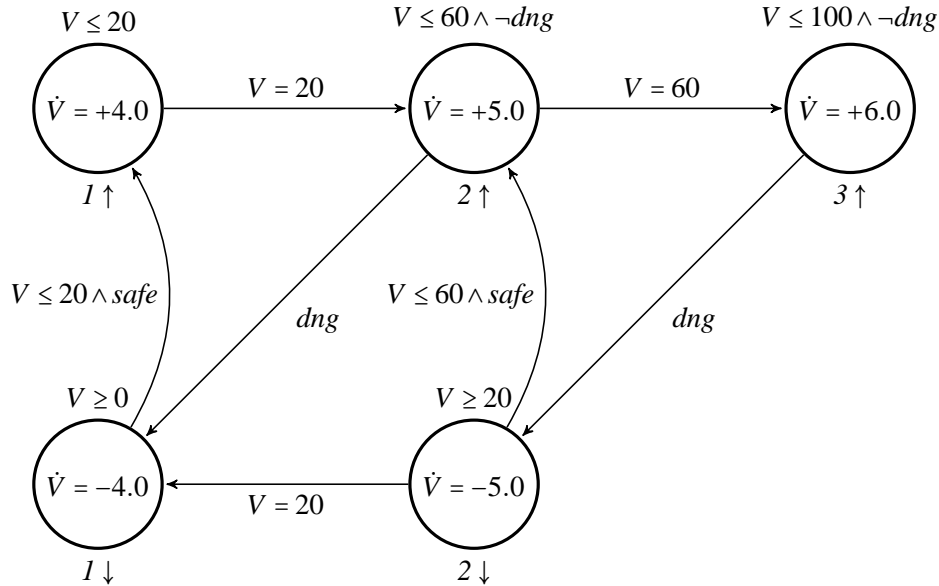

\begin{figure}
{\scriptsize
\begin{center}
\begin{align*}
    &\init \clauseif \exists\ \mathit{V,G,WG} \big(
    \atell{G=[1\uparrow\mid\_]} \parallel
    \achange{V}{0}{+4.0} \parallel
    \atell {V\geq 0 \wedge V\leq 100} \parallel
    \texttt{gearbox}(\mathit{G, WG, V}) \parallel
    \texttt{watcher}(\mathit{WG}) \big)\\ 
    &\texttt{gearbox}(\mathit{G,WG,V}) \clauseif \exists\ \mathit{G',WG'} \Big(\\
    &\begin{aligned}[t]
        &\aaskc{G=[1\uparrow\mid\_] \wedge V \leq 20}
        +\aaskc{G=[2\uparrow\mid\_] \wedge V \leq 60 \wedge \mathit{WG}\neq[\mathit{dng}\mid\_]}
        +\aaskc{G=[3\uparrow\mid\_] \wedge V \leq 100 \wedge \mathit{WG}\neq[\mathit{dng}\mid\_]}\\
        +&\aaskc{G=[1\downarrow\mid\_] \wedge V \geq 0}
        +\aaskc{G=[2\downarrow\mid\_] \wedge V\geq 20}\\
        +&\aask{G=[1\uparrow\mid\_] \wedge V=20}\big(
        \atell{G =[1\uparrow\mid G']} \parallel
        \atell{G'=[2\uparrow\mid\_]} \parallel
        \achange{V}{\_}{+5.0} \parallel
        \texttt{gearbox}(\mathit{G', WG, V})\big)\\
        +&\aask{G=[2\uparrow\mid\_] \wedge V=60}\big(
        \atell{G =[2\uparrow\mid G']} \parallel
        \atell{G'=[3\uparrow\mid\_]} \parallel
        \achange{V}{\_}{+6.0} \parallel \texttt{gearbox}(\mathit{G', WG, V})\big)\\
        +&\aask{G=[2\downarrow\mid\_] \wedge \mathit{V}=20}\big(
        \atell{G=[2\downarrow\mid \mathit{G'}]} \parallel
        \atell{G'=[1\downarrow\mid\_]} \parallel
        \achange{V}{\_}{-4.0} \parallel
        \texttt{gearbox}(\mathit{G', WG, V})\big)\\
        +&\aask{G=[2\uparrow\mid\_] \wedge \mathit{WG}=[\mathit{dng}\mid\_]}\big(
        \begin{aligned}[t]
            &\atell{G =[2\uparrow\mid G']} \parallel
            \atell{G'=[1\downarrow\mid\_]} \parallel
            \atell{\mathit{WG}=[\mathit{dng}\mid \mathit{WG'}]} \parallel\\
            & \achange{V}{\_}{-4.0} \parallel
            \texttt{gearbox}(\mathit{G', WG', V})\big)
        \end{aligned}\\
        +&\aask{G =[3\uparrow\mid\_] \wedge \mathit{WG}=[\mathit{dng}\mid\_]}\big(
        \begin{aligned}[t]
            &\atell{G =[3\uparrow\mid G']} \parallel
            \atell{G'=[2\downarrow\mid\_]} \parallel
            \atell{\mathit{WG}=[\mathit{dng}\mid \mathit{WG'}]} \parallel\\
            &\achange{V}{\_}{-5.0} \parallel
            \texttt{gearbox}(\mathit{G', WG', V})\big)
        \end{aligned}\\
        +&\aask{G=[1\downarrow\mid\_] \wedge \mathit{WG}=[\mathit{safe}\mid\_] \wedge V\leq20}\big(
        \begin{aligned}[t]
            &\atell{G =[1\downarrow\mid G']} \parallel
            \atell{G'=[1\uparrow\mid\_]} \parallel
            \atell{\mathit{WG}=[\mathit{safe}\mid \mathit{WG'}]} \parallel\\
            &\achange{V}{\_}{+4.0} \parallel
            \texttt{gearbox}(\mathit{G', WG', V})\big)
        \end{aligned}\\
        +&\aask{G =[2\downarrow\mid\_] \wedge \mathit{WG}=[\mathit{safe}\mid\_] \wedge V\leq60}\big(
        \begin{aligned}[t]
            &\atell{G=[2\downarrow\mid G']} \parallel
            \atell{G'=[2\uparrow\mid\_]} \parallel
            \atell{\mathit{WG}=[\mathit{safe}\mid \mathit{WG'}]} \parallel\\
            &\achange{V}{\_}{+5.0} \parallel
            \texttt{gearbox}(\mathit{G', WG', V})\big)\Big)
        \end{aligned}\\
    \end{aligned}\\
    &\texttt{watcher}(\mathit{WG}) \clauseif \exists\ \mathit{WG'} \Big( 
    \begin{aligned}[t]
        &\aaskc{\mathit{true}}\\
        +& \aask{\mathit{true}}
        \big(\atell{\mathit{WG}{=}[\mathit{safe}\mid\mathit{WG'}]}
        \parallel \texttt{watcher}(\mathit{WG'})\big)\\
        +& \aask{\mathit{true}}
        \big(\atell{\mathit{WG}{=}[\mathit{dng}\mid\mathit{WG'}]}
        \parallel \texttt{watcher}(\mathit{WG'})\big)\Big)
    \end{aligned}\\
\end{align*}
\end{center}
\caption{\hytccp\ model for a gear shift system}
\label{fig:gearHytccp}
}
\end{figure}

\section{Related Work}
\label{sec:rel_work}
In \cite{GuptaJSB94}, \hcc{} was introduced
as the first extension over continuous time of the concurrent
constraint paradigm.
Although both \hytccp{} and \hcc{} are declarative languages with a logical nature,
there are some important differences between them.
First of all, \hytccp{} is a non-deterministic language, while \hcc{} is deterministic.
We believe that this is an essential feature
for modeling hybrid systems, which are inherently non-deterministic.
\hytccp{} has been defined as a modeling language for hybrid systems in the
style of hybrid automata. This means that we aim to obtain
programs with a structure similar to that of hybrid automata,
but described in a more abstract way.
The non-deterministic choice is a powerful construct
that allows the set of all possible transitions of an
hybrid automata to be expressed as a list of $\aask{}$ and $\aaskc{}$ branches.
Furthermore, in \hcc{}, the information on the value and flow of continuous variables is modeled
as a constraint of the underlying continuous constraint system.
On the contrary, in \hytccp{}, there is a clear distinction between discrete and continuous variables.
In \hcc{} the positive information in the store must
be transferred by using the agent $\taghence$.
In contrast, in \hytccp{} the positive information
in the discrete store is transferred automatically from one step to the next.

In \cite{BeekMRRS06} and \cite{CuijpersR05},
two process algebras for hybrid systems
have been defined: \textit{Hybrid Chi} and \textit{HyPa}, \resp.
The process algebra \textit{Hybrid Chi} \cite{BeekMRRS06} shares with \hytccp{}
the separation between
discrete and continuous variables, the synchronous nature and
the concept of delayable guard (corresponding to the suspension of the non-deterministic choice).
\textit{HyPa} \cite{CuijpersR05} was introduced as
an extension of the process algebra \textit{ACP}.
It differs from \textit{Hybrid Chi}
mainly in the way time-determinism is treated, and in the modeling of time passing.

\section{Conclusions}
\label{sec:conclusions}
In this paper we have presented \hytccp{}, an extension of \tccp{}
over continuous time with the aim of modeling hybrid systems
in a declarative and logical way by
abstracting away from all the implementation details.
\hytccp{} has been introduced as a
synchronous and non-deterministic language defining computations similar to that of hybrid automata.

\hytccp{} has several advantages that make it suitable for modeling hybrid systems.
Its declarative nature facilitates
a high level description close to that of hybrid automata.
In addition, the logical nature of \hytccp{}
eases the development of formal methods techniques
for the static analysis and verification of hybrid systems.
Furthermore, since \hytccp{} is a conservative extension of \tccp{},
it is possible to describe with the same syntax
concurrent, reactive and hybrid systems.

In the future, we plan to
develop a framework for the description and simulation of
\hytccp{} programs. In this way, we will be able
to model complex hybrid systems in \hytccp.
Given the affinity of the two formalisms,
we are interested in defining
a translation rules system from \hytccp{} to hybrid automata and viceversa,
in order to transfer verification and analysis results from one formalism to the other.
Furthermore, we plan to use model checking and abstract interpretation techniques to
verify temporal properties of hybrid systems written in \hytccp{} (as done in \cite{GallardoP13} for SPIN
and in \cite{CominiTV14cltl} for \tccp).
Another feature we would like to explore is the adjustment of the language to
make it compatible with rectangular hybrid automata \cite{Kopke96}.

\bibliographystyle{eptcs}
\bibliography{biblio-laura}

\end{document}